\documentstyle [a4,12pt]{article}
\newcommand{\mincir}{\raise -2.truept\hbox{\rlap{\hbox{$\sim$}}\raise5.truept
\hbox{$<$}\ }}
\newcommand{\magcir}{\raise -2.truept\hbox{\rlap{\hbox{$\sim$}}\raise5.truept
\hbox{$>$}\ }}
\newcommand{\minmag}{\raise-2.truept\hbox{\rlap{\hbox{$<$}}\raise 6.truept\hbox
{$>$}\ }}
\newcommand{\be}{\begin{equation}}
\newcommand{\ee}{\end{equation}}
\newcommand{\ba}{\begin{eqnarray}}
\newcommand{\ea}{\end{eqnarray}}
\newcommand{\brr}{\begin{array}}
\newcommand{\err}{\end{array}}
\newcommand{\bc}{\begin{center}}
\newcommand{\ec}{\end{center}}

\newcommand{\lb}{{\left<\right.}}
\newcommand{\rb}{{\left.\right>}}
\newcommand{\hm}{\,h^{-1}{\rm Mpc}}

\title{ {\bf Large--Scale Angular Correlations} \\
{\bf in CDM Models}}
\author{
{\bf Lauro Moscardini}$^1$, {\bf Stefano Borgani}$^2$, \\
{\bf Peter Coles}$^3$, {\bf Francesco Lucchin}$^1$, \\
{\bf Sabino Matarrese}$^4$,
{\bf Antonio Messina}$^5$
\& {\bf Manolis Plionis}$^6$ \\ ~\\
{\em $^1$Dipartimento di Astronomia, Universit\`a di Padova,} \\
{\em vicolo dell'Osservatorio 5, I--35122 Padova, Italy} \\ ~\\
{\em $^2$INFN, Sezione di Perugia,}\\
{\em c/o Dipartimento di Fisica dell'Universit\`a,} \\
{\em via A. Pascoli, I--06100 Perugia, Italy} \\ ~\\
{\em $^3$Astronomy Unit, School of Mathematical Sciences,} \\
{\em Queen Mary \& Westfield College, Mile End Road,} \\
{\em London, E1 4NS, UK} \\ ~\\
{\em $^4$Dipartimento di Fisica G. Galilei, Universit\`a di Padova,} \\
{\em via Marzolo 8, I--35131 Padova, Italy} \\ ~\\
{\em $^5$Dipartimento di Fisica A. Righi, Universit\`a di Bologna,} \\
{\em via Irnerio 46, I--40126 Bologna, Italy} \\ ~\\
{\em $^6$SISSA -- International School for Advanced Studies,} \\
{\em via Beirut 2--4, I--34014 Trieste, Italy} \\ ~\\
{\sl The Astrophysical Journal Letters, submitted} }
\date{}
\begin{document}

\maketitle

\newpage
\section*{Abstract}
We investigate the behaviour of the angular (projected) galaxy--galaxy
correlation function, $w(\vartheta)$, in the framework of
Cold Dark Matter (CDM) models. We compare the situation for the standard
CDM model with Gaussian (i.e. random--phase) initial fluctuations with
comparable CDM models with non--random phases. To do this, we have generated
artificial Lick maps using $N$--body simulations so that we reproduce the
main features of this catalog as accurately as we can.
We compare the $w(\vartheta)$ measured from the simulations with
the APM correlation (scaled to the depth of the Lick map).
For the Gaussian CDM model, we find that neither the standard normalisation
($b=1.5$) nor a more evolved model ($b=1$) (as suggested by the COBE data),
can reproduce the correlations on large angular scales ($\vartheta
{}~\magcir 2.5^\circ$). We come to a similar conclusion about
CDM models with positively skewed initial fluctuation distributions,
and can therefore exclude this choice for initial non--random phases.
In contrast, models with initially negatively skewed fluctuations
produce a $w(\vartheta)$ that declines much more gently on large scales.
Such models are therefore, at least in principle,
capable of reconciling the lack of large--scale
power of the CDM spectrum with the observed clustering of APM galaxies.

\vspace{0.5cm}
\noindent{\em Subject headings:} Galaxies: formation, clustering -- large-scale
structure of the Universe -- early Universe -- dark matter.
\vspace{0.5cm}

\section {Introduction}
The Standard Cold Dark Matter (SCDM) model has long been the `standard'
model of galaxy formation and clustering because of its great success at
explaining the small scale dynamics and clustering of galaxies, once these
are identified with high peaks of the initial density fluctuations, in the
spirit of the {\em biased} galaxy formation model (see Frenk 1991 for a
review). In recent times, however, the availability of various data sets,
which are sensitive to larger scale features of the galaxy distribution, has
caused a re--evaluation of the status of the SCDM picture.

A variety of different observational
probes of the large--scale distribution of matter
in the Universe indicate that the great weakness of SCDM is that
it lacks sufficient power on scales $> 20-30\hm$ to be consistent
with the real Universe. The first compelling evidence that this was the
case came with the measurement of the angular two--point
correlation function of galaxies identified in projection on the
sky using the APM device (Maddox et al. 1990$b$). The great advantage
of these data over previous analyses of the Lick catalog (Groth \& Peebles
1977) is the careful
control of systematic errors achieved by using an automated plate--scanning
device (Maddox et al. 1990$a,c$). Indeed, an analysis of
angular correlations of galaxies identified using the COSMOS machine
(Heydon--Dumbleton, Collins, \& MacGillivray 1989) has come to similar
conclusions to those of APM (Collins, Nichol, \& Lumsden 1992).
Nevertheless, there does remain some residual doubt that there may be
systematic errors in the calibration of the magnitude limit on the
plates scanned by these automatic machines (Fong, Hale--Sutton, \& Shanks
1992). For such errors to be responsible for the excess power detected
in $w(\vartheta)$ requires them to be correlated amongst adjacent plates
and, while this is not impossible, no compelling mechanism has been
suggested as to how these errors might be introduced.

Subsequent analyses of the three--dimensional clustering of IRAS galaxies in
the QDOT catalog ({\em e.g.}, Efstathiou et al. 1990; Saunders et al. 1991)
and, more recently,
of a redshift survey of APM galaxies (Loveday et al. 1992) provide
independent confirmation of the projected APM \& COSMOS results.
Moreover, the
temperature fluctuations in the Cosmic Microwave Background (CMB) detected
recently by the COBE team (Smoot et al. 1992) indicate a higher
fluctuation amplitude on large scales than expected in SCDM. There is also
evidence for higher fluctuations in the mass distribution from galaxy
peculiar motions (Bertschinger et al.
1990) and the large--scale clustering of galaxy clusters (Batuski, Melott,
\& Burns 1987).

Various possible remedies for this large--scale weakness have been
suggested in the literature. A direct method of obtaining extra
fluctuations on large scales is to introduce a dark matter component with a
larger coherence length than CDM. The resulting `mixed dark matter'
(MDM) models, in which there is both a cold and a hot component (Valdarnini
\& Bonometto 1985; Achilli, Occhionero, \& Scaramella 1985), seem to agree
with most of the clustering observations (van Dalen \& Schaefer 1992;
Taylor \& Rowan--Robinson 1992; Davis, Summers, \& Schlegel 1992; Klypin et
al. 1992). A low--density CDM model with $\Omega_0 \sim 0.2$ is also a
viable possibility, especially if one adds a cosmological constant term so
that the resulting model is spatially flat (Efstathiou, Sutherland, \&
Maddox 1990). It has been speculated that a CDM model with a lower bias
might be in accord with the observations (Couchman \& Carlberg 1992); more
complex astrophysical effects may also induce a scale--dependent bias
(Babul \& White 1991; Bower et al. 1993; Coles 1993). Choosing a primordial
fluctuation spectrum which is tilted away from the standard Zel'dovich
scale--invariant form (Lucchin \& Matarrese 1985; Salopek, Bond, \& Bardeen
1989; Adams et al. 1993) is another way to introduce large--scale power,
but this alternative is strongly constrained (Vittorio, Matarrese, \&
Lucchin 1988; Liddle, Lyth, \& Sutherland 1992; Cen et al. 1992; Tormen et
al. 1993). Another alternative is to invoke a skewed (i.e., non--Gaussian)
distribution of primordial fluctuations whilst keeping the primordial
scale--invariant spectrum (Moscardini et al. 1991, hereafter MMLM;
Matarrese et al. 1991; Messina et al. 1992; Weinberg \& Cole 1992). All
these possibilities lack the compelling theoretical simplicity of the SCDM
model but we must consider all of them as potentially viable until excluded
by empirical data.

Given the prominence of the APM \& COSMOS angular correlation functions
amongst the evidence against CDM, we decided to look in detail at the
behaviour of the angular correlations of galaxies in CDM models. In this
{\em Letter} we shall address two main questions.

First is the question as to what extent the APM result actually rules
out SCDM, or even CDM with a low bias \`{a} la Couchman \& Carlberg
(1992). Previous analyses of this question have used a mixture of
linear theory and the Limber equation to compare the expected angular
correlation function with the observed one.
Fortunately we have already extracted realistic projected galaxy catalogs
from full $N$--body simulations with CDM power--spectrum (Coles et al. 1993$b$,
hereafter CMPLMM; Borgani et al. 1993).
We can therefore perform a much more
direct evaluation of the expected correlation function from numerical
simulations than is possible with the usual mixture of analytic
and numerical techniques. In Section 2 below,
we shall therefore use these simulations to
test whether SCDM, or CDM with a lower bias, is truly at variance
with the data.

The second question is whether any of the alternatives listed
above are also excluded by angular correlation data. In the course
of a study of the topology of the large--scale distribution of galaxies,
CMPLMM generated simulated projected catalogs
for CDM models with skewed initial fluctuation statistics. We shall
therefore, in Section 3,
take this opportunity to investigate whether skewed CDM models
can be excluded by the data.

\section {Standard CDM model}
The first model we want to study is the SCDM model, characterized by
the initial density fluctuation spectrum ${\cal P}(k) \propto k ~T^2(k)$,
with $T(k)$ the transfer function appropriate for CDM (e.g. Davis et al.
1985). We consider a flat universe model with vanishing cosmological constant
and Hubble constant $h=0.5$ in units of $100$ km sec$^{-1}$ Mpc$^{-1}$.

Our analysis is based on $N$--body simulations, described in detail
by Messina et al. (1992) and Lucchin et al. (1993),
which use a PM code with $N_p=128^3$ particles, $N_g=128^3$ grid--points on
a cubic box of side $L=260 ~h^{-1}$ Mpc.
The `present time' is fixed so that the variance of linear mass--fluctuations
in a sphere of radius $8~h^{-1}$ Mpc, $\sigma^2_8$, is unity; this
normalization is
consistent with the recent COBE detection of large scale CMB
anisotropies (Smoot et al. 1992).
We also show results for a biased model with linear bias parameter
$b=1.5$ (i.e. we consider the same simulation, but at the time
when $b \equiv \sigma_8^{-1}=1.5$), previously referred as SCDM.
With the latter normalization
the slope of the galaxy spatial correlation function
is best fitted by the observed value $\gamma=1.8$.
Note, however, that this normalization does not agree with
the level of CMB fluctuations detected by COBE; moreover, gravitational
waves do not reach
a level suitable to fill the gap in this particular case
(Davis et al. 1992; Liddle \& Lyth 1992; Lidsey \& Coles 1992;
Lucchin, Matarrese, \& Mollerach 1992; Salopek 1992; Souradeep \& Sahni 1993).

In order to have a direct determination of the angular correlation function
$w(\vartheta)$ in the simulations, we use our three--dimensional data to
construct artificial catalogs in projection. As in CMPLMM, we want to mimic
the Lick map. As discussed in detail in
that paper, the construction of mock Lick maps requires quite a large galaxy
number density, $3 \times 10^{-2}~h^3$ Mpc$^{-3}$, corresponding to
$530,000$ galaxies in the whole simulation box.
To select these galaxies in our low--resolution simulations we
smooth the initial density field with a Gaussian
filter of radius $1~h^{-1}$ Mpc and pick up all particles in regions above
a density threshold, fixed in such a way that the object number density
equals the required value.
Due to the rather simplified galaxy identification procedure,
we can only assume that our selected objects roughly trace the actual galaxy
distribution.

To build up our projected Lick look--a--like catalogs we then proceed
as follows. Given the box--side (260 $h^{-1}$ Mpc) and the solid angle
we want to study ($b^{\mbox{\tiny II}} \ge 45^{\circ}$),
we need to replicate the original simulation box exploiting
its periodic boundaries. In fact, although
the characteristic depth of the Lick map is only $D^{*} \sim 210$ $h^{-1}$ Mpc
(Groth \& Peebles 1977), galaxies with much larger distances
may also enter the catalog. The precise way this replication is done
is described in CMPLMM: it requires three levels of replicated boxes for
an overall number of 56 ones.
We then assign to each of the
$\sim 530,000$ galaxies an absolute magnitude according to the Schechter
(1976) luminosity function,
$\Phi(M) \sim \mbox{dex } [-0.4(\alpha+1)M] \exp [-\mbox{dex } 0.4(M^{*}-M)]$,
with $\alpha=-1.26$, $M^{*}=-19.6$,
truncated at both faint and bright ends so that $dN/dM = 0$ for
$M > M^{*}+3$ and $M < M^{*}-2$; next, we determine the apparent magnitude,
corresponding to its distance, taking also into account $K$--corrections
and expansion effects. Finally, we select galaxies with the same magnitude
limit as in the Lick map ($m_{lim} \le 18.8$).
CMPLMM verified the robustness of the results to variations of the
luminosity function and contamination due to replication of the original box.
The resulting projected galaxy distribution is then binned in 10$\times 10$
arcmin cells, in the same way as provided by the Lick catalog.

In order to evaluate the angular two--point correlation function,
$w(\vartheta)$, we use the estimator
\be
w(\vartheta)~=~{\lb n_in_j\rb _\vartheta
\over \lb {1\over 2}(n_i+n_j)\rb _\vartheta^2} -1\,.
\label{eq:wth}
\ee
In eq.(1), $n_i$ and $n_j$ are the galaxy counts in
the $i$--th and $j$--th cell,
respectively, placed at separation $\vartheta$; $\lb \cdot
\rb _\vartheta$ indicates an average taken over all
cell pairs with separation $\vartheta$. Due to the huge number of cells,
which makes the $w(\vartheta)$ computation numerically expensive, we prefer
to adopt two different procedures at small and large angular scales. For
$\vartheta \le 2^\circ$ we collect 10$\times 10$ arcmin cells in $6^\circ
\times 6^\circ$ plates. Then, the correlation function is evaluated within
each of these plates and the results are averaged to give the final
$w(\vartheta)$. At larger scales the smaller cells are grouped to form
counts in $1^\circ \times 1^\circ$ cells, and $w(\vartheta)$ is evaluated
according to eq.(\ref{eq:wth}), with $n_i$ and $n_j$ provided by the counts
in such larger cells. The reliability of our method is confirmed by the
smooth shape of $w(\vartheta)$ around $\vartheta \simeq 2^\circ$. The
results of the analysis for the Gaussian CDM models are shown in Figure 1.
We also plot the APM data about $w(\vartheta)$, scaled to the depth of
the Lick map, as described by Maddox et al. (1990$b$).

The lack of power on scales larger than $\vartheta \simeq 2.5^\circ$ of the
SCDM model (with $b=1.5$) is clearly evident in Figure 1; the same is also
true for the more evolved ($b=1$) model. A low--bias ($b~\mincir 1$) model
has been advocated by Couchman \& Carlberg (1992) in order to alleviate the
problems of CDM at scales $\sim 20-40\hm$. These authors claim that a more
evolved CDM distribution than the standard normalisation would give a
galaxy distribution consistent with the APM data. On the contrary, we find
that the only effect of further evolving the CDM spectrum is that of
increasing the small--scale correlation amplitude, while leaving the
large--scale tail substantially unchanged. It is clear that a more detailed
comparison between our results and that of Couchman \& Carlberg (1992)
would be difficult to realize. In fact, their method to estimate the
angular function is rather indirect and based on projecting, via the Limber
equation, the correlation data extracted from the three--dimensional
simulation box. Instead, in our analysis we tried to reproduce as
accurately as we can the observational setup relevant for the Lick catalog;
assign luminosities to galaxies, project them in an observational cone,
define a magnitude--limited sample and perform the analysis directly on the
angular distribution. Although the method of identifying galaxies in our
simulations is also different, we do not expect this to play an important
role in determining the results.

\section {Skewed CDM models}
An interesting alternative to the SCDM model
is provided by the possibility that primordial perturbations
had non--random phases. These could result either from suitable
inflationary models or in more specific models based on phase transitions
in the early universe.
CDM models with non--Gaussian initial conditions have been considered by
MMLM (see also Matarrese et al. 1991; Messina et al. 1992), who
have shown that both the clustering dynamics and the present
large--scale structure of the universe
are strongly affected by the sign of the primordial skewness of mass
fluctuations. Models with a positive skewness rapidly cluster to a lumpy
structure with small coherence length, while negative skewness models build
up a cellular structure
by the slow process of merging of shells around primordial underdense
regions, with larger coherence length. Among
these non--Gaussian models the skew--negative ones appear more successful
at reproducing the observed properties of the large--scale structure in the
framework of CDM models. Similar
results are also obtained by Weinberg \& Cole (1992), who start from
scale--free skewed non--Gaussian models. CMPLMM have analyzed the
two--dimensional topology and found that both Gaussian and skew--positive
models do not fit observations, while skew--negative ones provides a much
better agreement. Similar results have also been found by Borgani et al.
(1993), who selected cluster samples from simulated Lick maps. After
applying a list of statistical tests to the resulting cluster distributions,
they found that several clustering features crucially depend on the initial
amount of phase correlation. Coles et al. (1993$a$) have shown that
skew--negative CDM models are much better at accounting for the observed
skewness in the distribution of QDOT galaxies.

The non--Gaussian CDM models we consider are the same analyzed by MMLM,
namely {\em Lognormal} (LN) and {\em Chi--squared} with one degree
of freedom ($\chi^2$), chosen as distributions for the peculiar
gravitational potential, $\Phi$, before the action of the CDM transfer
function. Each non--Gaussian distribution actually splits in two models:
the {\em positive} ($LN_p$ and $\chi^2_p$) and {\em negative} ($LN_n$ and
$\chi^2_n$) ones, according to the sign of the skewness for linear
mass--fluctuations. Skew--positive/negative models are characterized
by a primordial excess of over/under--dense regions, which is at the
origin of their dynamical behaviour. Initial conditions are assigned in
terms of the peculiar gravitational potential. This can be obtained
in Fourier space as $\tilde{\Phi}({\bf k}) = T(k)\, F(k)\,
\tilde{\varphi}({\bf k})$, where $\varphi$ is a non--Gaussian random
field, related to a standard Gaussian one $w$, with flicker--noise
spectrum, by a local, non--linear map: $\varphi({\bf x}) \propto e^{w({\bf
x})}$ for $LN$ and $ \varphi({\bf x}) \propto w^2({\bf x})$ for $\chi^2$. A
smooth correction factor $F(k)$ is also applied so that all our models
start with exactly the standard CDM power--spectrum.

Projected maps are obtained according to the procedure described above.
In Figure 2 the angular correlation function $w(\vartheta)$ is shown both
for skew--positive and skew--negative models at the time when $b=1$.
As for the skew--positive models, there are no appreciable differences
with respect to the Gaussian case. This is quite easy to understand, since
for these models the effect of the initial non--random phases is to add
coherence at small scales, where subsequent non--linear gravitational
evolution re--arranges the clustering.

Much more interesting is the behaviour of skew--negative models. The
large--scale coherence they introduce has the effect of substantially
increasing the correlation amplitude at $\vartheta~\magcir 2^\circ$. From the
right panel in Figure 2 it is evident that the $LN_n$ model gives rise to an
exceeding amount of large--scale power, which agrees with the presence of
huge coherent structures generated by this model even after projection
(see Figure 2 of CMPLMM).
On the other hand, the $\chi^2_n$ model produces a rather adequate
large--scale correlation, although the $w(\vartheta)$ slope is still
slightly flatter than observed. However, in this paper we are not searching
for a best--fit non--Gaussian model. Instead, we
are investigating the effect of introducing phase correlations in the
initial conditions. In this spirit, Figure 2 suggests that a
skew--negative model with a rather limited non--Gaussian behavior
(such as the $\chi^2_n$ model) succeeds at accounting for the large--scale
power displayed by the APM data, even within the CDM scenario.

A further question concerns whether changing the definition of present time
in the simulations leads to substantially different results. To answer this,
we also decided to identify the present when the slope of the spatial
two--point correlation function for galaxies matches the observed one. This
epoch corresponds to $b \approx 2$ for the skew--positive models and $b
\approx 0.5$ for the skew--negative ones. The results for these cases are
not reported here, since in both cases $w(\vartheta)$ corresponds to a
worse fit: skew--positive models, being less evolved, have less power on all
scales; skew--negative ones, being more evolved, have exceedingly high power.

\section {Conclusions}
We have obtained two important results in this paper which allow us
to answer the two questions we posed in the introduction.

First, and contrary to the suggestion of Couchman \& Carlberg (1992),
we find that CDM models with Gaussian initial perturbations cannot
reproduce the APM correlation function, even with a higher normalisation than
the standard scenario.

Second, and consistent with results we have obtained in other papers,
we have demonstrated that it is possible to reconcile the CDM hypothesis
with observations of galaxy clustering and CMB temperature
anisotropies (pointing toward a low--bias, $b\simeq 1$, choice) by introducing
non--Gaussian primordial fluctuations.

We have found the form of $w(\vartheta)$ to be very sensitive to the presence
of initial phase correlations.
In particular, models with positive skewness look rather similar to the
Gaussian ones; they introduce phase correlations only at small scales,
where non--linear gravitational evolution subsequently re--arrange the
clustering, while leaving unchanged the large--scale pattern.
The skew--negative models we have
looked at generate lots of power on large scales; we can even exclude one
of these models, $LN_{n}$, because it produces too high a clustering
amplitude on large scales.
The model that fits the APM correlation function
best is the $\chi^{2}_{n}$ model which generates a reasonable amount of power
on large angular scales, even with $b=1$.

The reason for the success of the skew--negative models compared to the
Gaussian case is the qualitatively different way in which perturbations
grow. Hierarchical clustering still acts on small scales, as in the Gaussian
CDM model, but large--scale structure grows by the
slow merging of shells around primordial underdensities to produce a
distribution with a very large coherence length. In this sense the behaviour
of these models resembles that of MDM ones, where small-- and
large--scale structures originate in a different way.

Since we have covered only an infinitesimal part of the space of all
skew--negative models, it is not surprising that we have not found one that
fits the data exactly, but we have shown that, at least qualitatively, such
models can explain the clustering data without too much difficulty.

Of course, we are not suggesting the $\chi^{2}_{n}$ as a physically
motivated model to fit to the observations. What we have done is
demonstrate that the APM data are not in contradiction with CDM
{\em per se}, merely with CDM {\em plus} the assumption of primordial
random phases. A CDM model with negatively skewed primordial statistics
is at least as successful as any other model
at explaining galaxy clustering data. We should take non--Gaussian models
seriously until they are definitely excluded by the data.

\newpage
\section* {Acknowledgments}
We are all very grateful to Steve Maddox for providing us with the APM
correlation function in digital form. SB wishes to acknowledge SISSA in
Trieste for the hospitality during several phases of preparation of this work.
PC thanks the Dipartimento di Astronomia at the Universit\`a di Padova for the
hospitality during a visit when some of this work was done. He also
acknowledges support from SERC under the QMW rolling grant GR/H09454. This
work has been partially supported by Ministero dell'Universit\`a
e della Ricerca Scientifica e Tecnologica and by Consiglio Nazionale
delle Ricerche (Progetto Finalizzato: Sistemi Informatici e Calcolo
Parallelo). The staff and the management of the CINECA Computer Center
(Bologna) are warmly acknowledged for their assistance and for allowing the
use of computational facilities.

\newpage
\large
\begin{center}
\noindent {\bf References}
\end{center}
\normalsize

\begin{trivlist}
\item[] Achilli, S., Occhionero, F., \& Scaramella, R. 1992, ApJ, 299, 577
\item[] Adams, F.C., Bond, J.R., Freese, K., Frieman, J.A., \& Olinto, A.V.
1993, Phys. Rev., D47, 426
\item[] Babul, A., \& White, S.D.M. 1991, MNRAS, 253, 31P
\item[] Batuski, D.J., Melott, A.L., \& Burns, J.O. 1987, ApJ, 322, 48
\item[] Bertschinger, E., Dekel, A., Faber, S.M., Dressler, A., \& Burstein,
D. 1990, ApJ, 364, 370
\item[] Borgani, S., Coles, P., Moscardini, L., \& Plionis, M. 1993, preprint
\item[] Bower, R.G., Coles P., Frenk C.S., \& White, S.D.M. 1993, ApJ, in press
\item[] Cen, R., Gnedin, N.Y., Kofman, L.A., \& Ostriker, J.P. 1992, ApJ, 399,
L11
\item[] Coles, P. 1993, MNRAS, in press
\item[] Coles, P., Moscardini, L., Lucchin, F., Matarrese, S., \& Messina, A.
1993$a$, MNRAS, submitted
\item[] Coles, P., Moscardini, L., Plionis, M., Lucchin, F., Matarrese, S.,
\& Messina, A. 1993$b$, MNRAS, 260, 572 (CMPLMM)
\item[] Collins, C.A., Nichol, R.C., \& Lumsden, S.L. 1992, MNRAS, 254, 295
\item[] Couchman, H.M.P., \& Carlberg, R.G. 1992, ApJ, 389, 453
\item[] Davis, M., Efstathiou, G., Frenk, C.S., \& White, S.D.M. 1985, ApJ,
292, 371
\item[] Davis, M., Summers, F.J., \& Schlegel, D. 1992, Nature, 359, 393
\item[] Davis, R.L., Hodges, H.M., Smoot, G.F., Steinhardt, P.J., \&
Turner, M.S. 1992, Phys. Rev. Lett., 69, 1856
\item[] Efstathiou, G., Kaiser, N., Saunders, W., Lawrence, A.,
Rowan--Robinson, M., Ellis, R.S., \& Frenk, C.S. 1990, MNRAS, 247, 10P
\item[] Efstathiou, G., Sutherland, W.J., \& Maddox, S.J. 1990, Nature,
348, 705
\item[] Fong, R., Hale--Sutton, D., \& Shanks, T. 1992, MNRAS, 257, 650
\item[] Frenk, C.S. 1991, in `Nobel Symposium No 79: The birth and early
evolution of our Universe', Physica Scripta, T36, 70
\item[] Groth, E.J., \& Peebles, P.J.E. 1977, ApJ, 217, 385
\item[] Heydon--Dumbleton, N.H., Collins, C.A., \& MacGillivray, H.T. 1989,
MNRAS, 238, 379
\item[] Klypin, A., Holtzman, J., Primack, J., \& Reg\H{o}s, E. 1992, preprint
\item[] Liddle, A.R., \& Lyth, D.H. 1992, Phys. Lett., B291, 391
\item[] Liddle, A.R., Lyth, D.H., \& Sutherland, W. 1992, Phys. Lett., B279,
244
\item[] Lidsey, J.E., \& Coles, P. 1992, MNRAS, 258, 57P
\item[] Loveday, J., Efstathiou, G., Peterson, B.A., \& Maddox, S.J. 1992,
ApJ, 400, L43
\item[] Lucchin, F., \& Matarrese, S. 1985, Phys. Rev., D32, 1316
\item[] Lucchin, F., Matarrese, S., Messina, A., \& Moscardini, L. 1993, in
Proc. of the International School of Physics `E. Fermi' on `Galaxy Formation',
eds. J. Silk and N. Vittorio, in press
\item[] Lucchin, F., Matarrese, S., \& Mollerach, S. 1992, ApJ, 401, L49
\item[] Maddox, S.J., Efstathiou, G., \& Sutherland, W.J. 1990$a$,
MNRAS, 246, 433
\item[] Maddox, S.J., Efstathiou, G., Sutherland, W.J., \& Loveday, J.
1990$b$, MNRAS, 242, 43P
\item[] Maddox, S.J., Sutherland, W.J., Efstathiou, G., \& Loveday, J.
1990$c$, MNRAS, 243, 692
\item[] Matarrese, S., Lucchin, F., Messina, A., \& Moscardini, L. 1991,
MNRAS, 252, 35
\item[] Messina, A., Lucchin, F., Matarrese, S., \& Moscardini, L. 1992,
Astroparticle Phys., 1, 99
\item[] Moscardini, L., Matarrese, S., Lucchin, F., \& Messina, A.
1991, MNRAS, 248, 424 (MMLM)
\item[] Salopek, D.S. 1992, Phys. Rev. Lett., 69, 3602
\item[] Salopek, D.S., Bond, J.R., \& Bardeen, J.M. 1989, Phys. Rev., D40, 1753
\item[] Saunders, W., et al. 1991, Nature, 349, 32
\item[] Schechter, P. 1976, ApJ, 203, 297
\item[] Smoot, G.F., et al. 1992, ApJ, 396, L1
\item[] Souradeep, T., \&  Sahni, V. 1993, MNRAS, in press
\item[] Taylor, A.N., \& Rowan--Robinson, M. 1992, Nature, 359, 396
\item[] Tormen, G., Moscardini, L., Lucchin, F., \& Matarrese, S. 1993, ApJ,
in press
\item[] Valdarnini, R., \& Bonometto, S.A. 1985, A\&A, 146, 235
\item[] van Dalen, A., \& Schaefer, R.K. 1992, ApJ, 398, 33
\item[] Vittorio, N., Matarrese, S., \& Lucchin, F. 1988, ApJ, 328, 69
\item[] Weinberg, D.H., \& Cole, S. 1992, MNRAS, 259, 652
\end{trivlist}
\newpage

\section*{\center Figure captions}

{\bf Figure 1.} The angular two--point correlation function, $w(\vartheta)$,
for the Gaussian CDM model at two evolutionary stages: $b=1.5$ (left panel)
and $b=1$ (right panel). Open squares and dashed lines are for the
simulated Lick maps, while filled dots are for the APM correlation, as
provided by Maddox et al. (1990$b$). It is apparent the lack of correlation at
$\vartheta ~\magcir 2.5^\circ$, which is not alleviated by leaving
the clustering evolving up to $b=1$.

\vspace{0.2truecm}
\noindent
{\bf Figure 2.} The same as in Figure 1, but for skewed CDM models, with
both positive (left panel) and negative (right panel) skewness. Only the
epoch associated to $b=1$ is considered. Open squares are for the
{\em Chi--squared} models, while open triangles are for the {\em Lognormal}
models. No significant differences with respect to the Gaussian case appears
for the skew--positive models. Negative skewness introduces
large--scale coherence, which, for the $LN_n$ model, gives an even exceeding
clustering at large separations. Instead, a rather adequate shape for
$w(\vartheta)$ is produced by the $\chi^2_n$ model.

\end{document}